\documentclass[aps,pre,showpacs,amsmath,amssymb,amsfonts,lengthcheck,twocolumn,longbibliography,superscriptaddress]{revtex4-2}
\usepackage{epsfig,graphicx,graphics,float}
\usepackage{xcolor}
\usepackage{amscd}
\usepackage{bm}
\usepackage{bbold}
\usepackage{psfrag}
\usepackage{mathrsfs}
\usepackage{soul}
\usepackage{bbm} 
\usepackage{chemformula}
\usepackage[caption=false]{subfig}
\usepackage{color}
\usepackage[bookmarks=true,colorlinks,linkcolor=blue,urlcolor=blue,citecolor=blue]{hyperref}

\usepackage{dutchcal}
\DeclareMathAlphabet{\matholdcal}{OMS}{cmsy}{m}{n}

\newcommand{\be}{\begin{equation}}
\newcommand{\ee}{\end{equation}}
\newcommand{\bea}{\begin{eqnarray}}
\newcommand{\eea}{\end{eqnarray}}

\newcommand{\mc}{\matholdcal}

\newcommand{\te}{\text}
\begin{document}

\title{Nanowelding of quantum spin-$\frac{1}{2}$ chains at minimal dissipation}

\author{Moallison F. Cavalcante}
\affiliation{Department of Physics, University of Maryland, Baltimore County, Baltimore, MD 21250, USA}
\affiliation{
Gleb Wataghin Physics Institute, The University of Campinas, 13083-859, Campinas, São Paulo, Brazil}
\author{ Marcus V. S. Bonan\c{c}a}
\affiliation{
Gleb Wataghin Physics Institute, The University of Campinas, 13083-859, Campinas, São Paulo, Brazil}
\author{Eduardo Miranda}
\affiliation{
Gleb Wataghin Physics Institute, The University of Campinas, 13083-859, Campinas, São Paulo, Brazil}
\author{Sebastian Deffner}
\affiliation{Department of Physics, University of Maryland, Baltimore County, Baltimore, MD 21250, USA}
\affiliation{National Quantum Laboratory, College Park, MD 20740, USA}
\date{\today}

\begin{abstract}
We consider the optimal control of switching on a coupling term between two quantum many-body systems. Specifically, we (i) quantify the energetic cost of establishing a weak junction between two quantum spin-$1/2$ chains in finite time $\tau$ and (ii) identify the energetically optimal protocol to realize it. For linear driving protocols, we find that for long times the excess (irreversible) work scales as $\tau^{-\eta}$, where $\eta=1, 2$ or a nonuniversal number depending on the phase of the chains. Interestingly, increasing a $J_z$ anisotropy in the chains suppresses the excess work thus promoting quasi-adiabaticity. The general optimal control problem is solved, employing a Chebyshev Ansatz. We find that the optimal control protocol is intimately sensitive to the chain phases.
\end{abstract}


\maketitle
\section{Introduction}
\label{introduction}

Recent developments in nano-engineering have taken great strides towards the realization of quantum technologies. Naturally, such \emph{quantum engineering} requires the delicate control of complex quantum many-body systems. This is complicated by the fact that in practice all processes have to occur at finite time, and hence parasitic nonequilibrium excitations are inevitable. Therefore, how to generate effectively adiabatic dynamics in genuinely nonequilibrium processes has received significant attention in the literature \cite{Polkovnikov_2008,RMP_Silva_2011,PRA_QST_2013,PRL_Bernier_2014,PRL_SD,Dora_1,Deffner2020EPL}. In addition, several experiments have been reported \cite{Chu2002,Jaksch_2005,Nature_Hofferberth_2007,Nature_Bason_2012,Science_Quant_simu_2017,NJP_Zhou_2018,PRX_control_2021} verifying the theoretical predictions.

Controlling complex many-body systems in finite time is described varying a Hamiltonian $\mc H(\lambda)$ according to an external control protocol $\lambda(t)$. If the rate of change of $\lambda(t)$ is larger than the ``gap'' the dynamics is intrinsically non-adiabatic \cite{sakurai2017modern}, leading to an accumulation of excess energy. Over the last decade, several techniques have been developed to minimize these undesired excitations, such as, for example, shortcuts to adiabaticity \cite{Berry_2009,PRL_del_campo_2012,Deffner2014PRX,AcconciaPRE2015,Deffner2016NJP,Claeys_2019,GueryRMP2019,roychowdhury2021entropy,Touil2021entropy} and optimal control strategies \cite{Thiago_1,Chamon_1,Marcus_2,Opt_LL,Soriani_2022}. Due to the complexity of the ensuing control fields, shortcuts to adiabaticity appear less practical in many-body systems. In the present work, we focus on optimal control strategies and therefore identifying such $\lambda(t)$ that minimize the excess work.

Among complex quantum systems, especially one-dimensional systems have received constant attention. This might be due to the various numerical and analytical techniques available, which allow for more controllable studies, or due to their unique physics \cite{giamarchi2004quantum}. Consider, e.g. a system of interacting electrons in a one-dimensional wire. In this case, Landau's Fermi liquid theory \cite{coleman2015introduction} breaks down, and hence the more sophisticated Luttinger liquid (LL) theory was developed. LL theory states that the low-energy excitations of a system of interacting electrons in $D=1$ are collective bosonic charge and spin excitations \cite{giamarchi2004quantum}. Moreover, LL theory is known to describe the low-energy sector of different interacting one-dimensional models, as the Hubbard model (away from half-filling) and the $XXZ$ Heisenberg chain in the critical phase \cite{gogolin2004bosonization}. In fact, the latter can be mapped (through a Jordan-Wigner transformation) onto a system of interacting spinless fermions, so LL theory is applicable to these systems as well. Finally, LL behavior has been experimentally verified in many systems \cite{Lake2005,PRL_exp_LL_2008,PRL_LL_exp_2017,Wu_2019,Wang_2022,LL_exp_2024}.   

The solid understanding achieved in these complex systems in $D=1$ has allowed for further investigations, such as, for instance, studies of junctions formed by coupling spin-chains or interacting nanowires, both in \cite{PRB_Kane_Fisher_1992,prb_Lal_2002,PRL_Egger_2002,prl_Chamon_2003,PRB_affleck,Nuclear_Cramp__2013,Y_junc_Buccheri,Nuclear_Giuliano_2020,PRB_jun_Domenico} and out of equilibrium \cite{PRB_Schiro_2015,join_Apollaro_2015,Landi,PRB_de_Paula_2017,PRE_join,PRA_cut_spin_chain}. The phenomena uncovered in such junctions exemplify notable themes in many-body physics, such as the multichannel Kondo physics \cite{PRL_Tsvelik_2013,Y_junc_Buccheri}, anomalous transport properties \cite{PRB_Kane_Fisher_1992,PRB_jun_Domenico} and exotic phases of matter in a network of quantum spin chains \cite{PRL_Ferraz_2019,PRB_Fabrizio_2024}. 

\begin{figure*}
    \centering
    \includegraphics[scale=0.18]{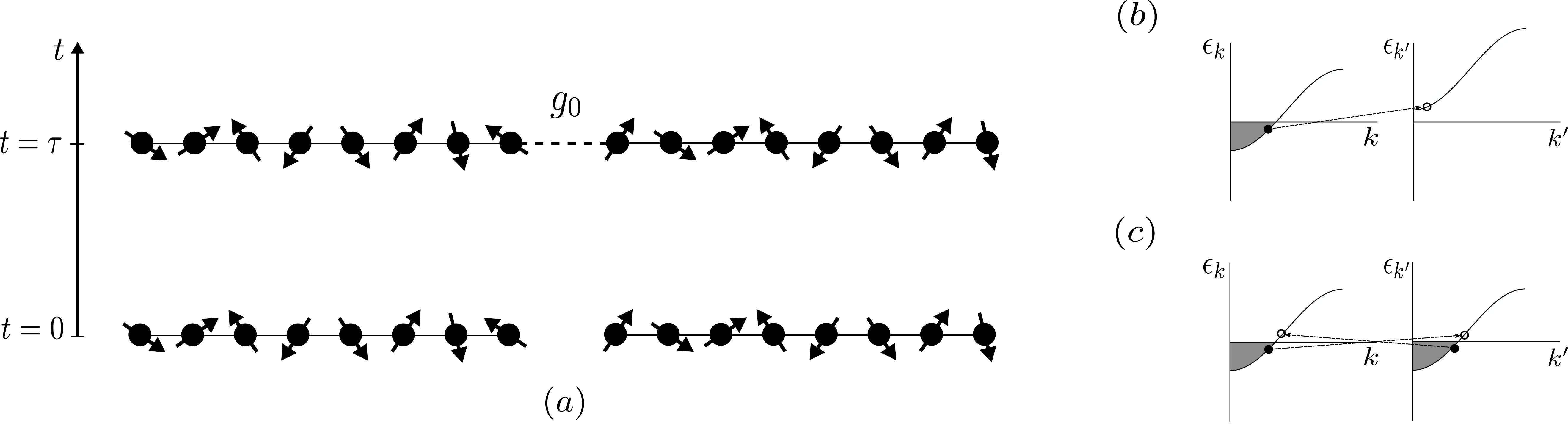}
    \caption{(a) Time-dependent nanowelding of two quantum spin-$1/2$ chains described by Eq. (\ref{junction}). The strength of the coupling between the chains increases from zero, at $t=0$, to a small final value $g_0$ at $t=\tau$. The most efficient protocol to realize such a junction is determined in Sec. \ref{sec_opt_prot} by requiring a minimal amount of energy spent during the process. (b) Scattering process between the two chains described by the lowest nonzero order in the perturbation theory considered here. The fermionic dispersion relation on the left represents the $XX$ gapless phase (the gray region denotes the Fermi sea), while on the right is the $XX$ gapped phase. (c) Same as (b) but for the case where both chains are in the $XX$ gapless phase.}
    \label{system}
\end{figure*}

Motivated by these studies, we consider the problem of quantifying the energetic cost and identifying the optimal strategy to establish, in finite time, a weak junction between two quantum spin-$1/2$ chains. We consider that for times $t<0$ two decoupled spin-$1/2$ chains are prepared in their respective ground state (not necessarily the same ground state). For times $t\geq 0$, the coupling between the end points of each chain is turned on according to a given time-dependent protocol until the boundary coupling strength reaches a small final value $g_0$ at a later time $\tau$. See Fig.~\ref{system} for a sketch of the physical scenario.

In the first part of the analysis, we consider processes in which the coupling strength between the chains is switched on through a linear ramp. In this case, we obtain analytical results for the energetic cost of establishing the junction when each chain is in one of two different, gapless or gapped, phases. For protocols of very short duration $\tau$, we find that the energetic cost scales as $\tau^2$ independent of the phase of the chains. For long times $\tau$, we find that, if there is no $J_z$-anisotropy in the chains, the energetic cost scales as a power-law $\tau^{-\eta}$ with a universal exponent $\eta$. This exponent is equal to $1$ when both chains are in the $XX$ gapless phase while it is equal to $2$ when one of the chains is in the $XX$ gapless phase but the other is in the $XX$ gapped phase. This last result corroborates what was shown in Ref.~\cite{Soriani_2022} but considering a more complicated system that is partially gapless. Remarkably, we have also shown that both of these scaling behaviors are \textit{protocol independent}, i.e., they do not depend on how the coupling strength increases in time (as long as $\tau$ is large). Once we include a $J_z$-anisotropy in the chains, specifically when both chains are in the critical LL phase, we find that the energetic cost scales as a power-law $\tau^{-\eta}$ but this time with a non-universal exponent $\eta=\eta(J_z)>0$ in the interval $(\sqrt{2})^{-1}<J_z<1$. This last result suggests that increasing interactions (here represented by $J_z$) can make the excess work decay faster with the process duration, hence promoting quasi-adiabaticity in finite time.

In the second part, we then determine the optimal control protocol, by requiring a minimal energetic cost to reach the final strength value $g_0$. The optimal protocol, when one of the chains is in the $XX$ gapped phase, is the same that was recently seen in the context of the transverse field Ising chain deep in the paramagnetic phase \cite{Soriani_2022}. When both chains are in the same $XX$ gapless phase, the optimal protocol is simply the linear protocol. When they are in the critical LL phase, however, the optimal protocol is an almost linear function of time, but differs slightly from it close to the initial $t=0$ and final $t=\tau$ points. 

As a main insight, our work demonstrates that the presence or absence of an energy gap and particle interactions are notable and important features in determining the excess work and the optimal control strategy.


\section{Model and quantum welding}
\label{sec_model}

We start by describing the model and establishing notions and notations. We consider two spin-$1/2$ chains of equal length $N$ described by the Hamiltonians,
\begin{eqnarray}
 H_{\ell=1,2}&=&J\sum_{j}\left(S^x_{j}S^x_{j+1} + S^y_{j}S^y_{j+1} + J_zS^z_{j}S^z_{j+1}\right)\nonumber\\
    &&+ \:h_{\ell}\sum_{j}S^z_{j},\label{H1}
\end{eqnarray}
where $S^{a=x,y,z}_j$ are spin-$1/2$ operators at site $j$, $J\equiv 1$ (our unit of energy) is the exchange coupling,  $J_z\geq0$ is the spin-anisotropy parameter, $h_{\ell}>0$ are the external magnetic fields in the $z$-direction of each chain, and the index $\ell=1,2$ denotes the two chains ($\ell=1$ for $-N\leq j\leq -1$ and $\ell=2$ for $1\leq j \leq N$). Notice that the only difference between the chains are the external magnetic fields $h_1$ and $h_2$. 

We now define the time-dependent protocol. For times $t<0$, the two spin-$1/2$ chains are decoupled and prepared in their respective ground states, so that the system state is $|\Upsilon_0\rangle=|\Upsilon_{1,0}\rangle\otimes|\Upsilon_{2,0}\rangle$, where $|\Upsilon_{\ell,0}\rangle$ is the ground state of chain $\ell$. For times $t\geq 0$, the coupling $\lambda(t)$ between them is turned on until a later time $\tau$; see Fig.~\ref{system} (a). Typically, the state $|\Upsilon(\tau)\rangle$ at time $\tau$ is rather complex due to the coupling between the degrees of freedom of the two chains. 

This process of ``welding'' the two chains is described by 
\begin{equation}
\mc H(\lambda)=H_0 + \lambda(t)V,
\label{junction}
\end{equation}
where $H_0=\sum_{\ell=1}^{2}H_{\ell}$ is the Hamiltonian of the separate chains and $V$ is the coupling term. This is a boundary interaction given by \cite{Affleck_1}
\begin{equation}
    V=g_0\left(S^{x}_{-1}S^{x}_{1} + S^{y}_{-1}S^{y}_{1} \right),
    \label{coupling}
\end{equation}
where $g_0\ll 1$ is a weak interaction strength. The protocol function $\lambda(t)$ that controls the joining process satisfies the boundary conditions $\lambda_i\equiv\lambda(0)=0$ and $\lambda_f\equiv\lambda(\tau)=1$, but we keep its time dependence generic.

It is important to point out that an extra term $\propto S^{z}_{-1}S^{z}_{1}$ could also appear in Eq. (\ref{coupling}). However, as we will discuss in Sec. \ref{sec_XXZ}, such a term has a scaling dimension larger than its $x,y$ counterparts shown in Eq.~(\ref{coupling}) \cite{Furusaki_1}. Therefore, from renormalization group arguments \cite{Vasseur2014}, the extra term goes to zero more rapidly than the term we keep and its effects are less relevant, which is why we do not include it.

\section{Energetic cost of welding}
\label{sec_ex}

To quantify the amount of energy that is spent during the finite-time welding protocol, we now evaluate the work done on the system, $W(\tau)$. Given that our system is isolated, this work is simply the difference between the final $\langle \mc H(\lambda_f)\rangle(\tau)$ and initial $\langle \mc H(\lambda_i)\rangle(0)$ energies \cite{Soriani_2022}, where $\langle\cdots\rangle(t)$ denotes the expectation value $\langle \Upsilon(t)|\cdots|\Upsilon(t)\rangle$. According to the Hellmann-Feynman theorem, $W(\tau)$ can be cast in the form
\begin{equation}
    W(\tau)=-\int_0^\tau dt\:\dot \lambda(t)\langle F(\lambda)\rangle(t),
    \label{work}
\end{equation}
where $\dot \lambda(t)$ is the time-derivative of $\lambda(t)$ and $F(\lambda)\equiv -\frac{\partial}{\partial \lambda}\mc H(\lambda)$ is the so-called generalized force. By construction we have $F(\lambda)=-V$.

In general, the work done, Eq.~\eqref{work}, encodes two distinct contributions \cite{Soriani_2022},
\begin{equation}
    W(\tau)=W_{\te{qs}}+W_{\te{ex}}(\tau).
\end{equation}
The first one, is the quasistatic contribution $W_{\te{qs}}$ which does not depend on the specific protocol, but only on the initial $\lambda_i$ and final $\lambda_f$ values. The second contribution, called the excess work $W_{\te{ex}}(\tau)$, is protocol-dependent and measures the nonadiabaticity of $W(\tau)$ given the finite-time protocol \cite{Jarzynski_2020}. In the limit $\tau \to \infty$, we expect to recover the adiabatic theorem \cite{sakurai2017modern,coleman2015introduction}, so $W_{\te{ex}}(\tau\to \infty)\to 0$ and, consequently, $W(\tau \to \infty)\to W_{\te{qs}}$.  

In the following, we are interested in determining (i) the scaling behavior of $W_{\te{ex}}(\tau)$ for large $\tau$ and (ii) the optimal driving protocol $\lambda_{\te{opt}}(t)$, i.e., the one that produces the minimal $W_{\te{ex}}(\tau)$, for particular, qualitatively different forms of $H_{1,2}$.


For a process that is coupled weakly enough, we can obtain a general expression for $W_{\te{ex}}(\tau)$.  In the interaction picture, the state of the system at time $\tau$ is given by \cite{bruus2004many}
\begin{equation}
    |\Upsilon(\tau)\rangle=e^{-iH_0\tau}\mc T\exp\left(-i\int_0^{\tau}dt\:\lambda(t)V_{I}(t)\right)|\Upsilon_0\rangle,
\end{equation}
where $\mc T$ is the time ordering operator and $V_{I}(t)=e^{iH_0t}Ve^{-iH_0t}$ is the coupling written in the interaction  picture, and we set $\hbar=1$. Using standard perturbation theory for weak couplings $g_0$ \cite{sakurai2017modern}, we expand the exponential and evaluate the expectation value of the generalized force~\eqref{work}. We obtain,
\begin{eqnarray}
    W_{\te{ex}}(\tau)&=&\frac{1}{2}\int_0^\tau dtdt'\:\dot \lambda(t)\dot\lambda(t')\Psi(t'-t) + \mc O(g^4_0),
    \label{excesswork}
\end{eqnarray}
which is nothing but the linear-response expression obtained in Ref.~\cite{Thiago_1}. As usual \cite{Kubo}, the relaxation function, $\Psi(t'-t)$, is defined in terms of the response function $\Phi(t'-t)$,
\begin{equation}
    \Psi(t'-t)=-\int_t^{t'}dt''\:\Phi(t''-t) +\te{const.},
    \label{relax_fun}
\end{equation}
where, for our case, 
\begin{equation}
    \Phi(t''-t)= i\langle [V_I(t''),V_I(t)]\rangle_0,
    \label{respon_fun}
\end{equation}
with $\langle\cdots\rangle_0\equiv\langle\Upsilon_0|\cdots |\Upsilon_0\rangle$. The relaxation function describes the evolution of the system towards equilibrium after an initial perturbation. In the derivation of Eq.~(\ref{excesswork}), we used the even parity of the relaxation function, $\Psi(s)=\Psi(-s)$, which is a consequence of the odd parity of the response function, $\Phi(s)=-\Phi(-s)$. The $\te{const.}$ in Eq.~(\ref{relax_fun}) guarantees that $\Psi(s\to \infty)\to 0$ \cite{Kubo}.

\paragraph*{Short-time behavior}


Before we proceed to a more complete analysis, it is interesting to discuss a general aspect of $W_{\te{ex}}(\tau)$ for weak processes with \textit{any} protocol. 
For processes of short duration, i.e., $\tau\ll 1$, or, more specifically, $\tau \ll \tau_R$, where $\tau_R$ is a system-dependent timescale which dictates the decay of $\Psi(s)$ \cite{Marcus_2}, we can determine $W_{\te{ex}}(\tau\ll \tau_R)$ universally.

The short-time behavior of the relaxation function can be obtained with help of the Baker-Hausdorff formula \cite{arfken2013mathematical}. Expanding $V_I(t)$ in powers of $t$ and evaluating Eq. (\ref{relax_fun}), we find 
\begin{equation}
    \Psi(s\ll \tau_R)\approx\te{const.} - \mc C s^2,
\end{equation}
where $\mc C\equiv\langle[[V,H_0],V] \rangle_0/2$ contains physical information about the system, $s=|t-t'|$ and $\te{const.}$ is the same constant appearing in Eq. (\ref{relax_fun}). 

Therefore, the excess work for any protocol takes the general form in very short-times
\begin{equation}
    W_{\te{ex}}(\tau\ll \tau_R)\propto \te{const.}' - \mc C \:\tau^2,
    \label{exc_work_short_tim}
\end{equation}
where $\te{const.}'$ is another nonuniversal constant. Then, as long as $\mc C\neq 0$, we conclude that $W_{\te{ex}}(\tau)$ scales, for very short-times, according Eq. (\ref{exc_work_short_tim}), independently of the details of the system (system details are encoded in $\mc C$ and $\te{const.}'$). The natural question arises, if similarly universal behavior can be found in the long-time behavior of $W_{\te{ex}}(\tau)$.

It is important to stress that up to this point we have not specified the protocol $\lambda(t)$. We now continue with linear protocols $\lambda(t)=t/\tau$, before returning to the general case in Sec.~\ref{sec_opt_prot}.

\section{Linear protocols}

\subsection{Case 1: $J_z=0$ -- the $XX$ model}
\label{sec_XX}

We begin by setting $J_z=0$. In this case, we can directly diagonalize $H_0$ by mapping the spins onto spinless fermions through the Jordan-Wigner (J-W) transformation \cite{Brezin1990,sachdev1999quantum}. Defining the notation, $S^a_{1,j}\equiv S^a_{j\leq -1}$ and $S^a_{2,j}\equiv S^a_{j\geq 1}$, the generalized J-W transformation used in \cite{Nuclear_Cramp__2013}, which accounts for the correct anti-commutation relations between fermions in different chains, reads
\begin{eqnarray}
    S^{+}_{1,j} &=& (-1)^j\chi_1\exp\left(i\pi\sum_{i=1}^{|j|-1}c^{\dagger}_{1, -i}c_{1, -i}\right)c^{\dagger}_{1, j},\nonumber
\\
    S^{+}_{2,j} &=& (-1)^j\chi_2\exp\left(i\pi\sum_{i=1}^{j-1}c^{\dagger}_{2, i}c_{2, i}\right)c^{\dagger}_{2, j},\nonumber
\\    
    S^z_{\ell,j} &=& c^{\dagger}_{\ell,j}c_{\ell,j}-1/2,
    \label{J_W_trans}
\end{eqnarray}
where $S^{\pm}_{\ell,j}=S^x_{\ell,j}\pm iS^y_{\ell,j}$ are the usual spin raising and lowering operators and $c_{\ell,j}$, with $\{c_{\ell,j},c^{\dagger}_{\ell',j'}\}=\delta_{\ell,\ell'}\delta_{j,j'}$, are spinless fermions. The Klein factors $\chi_{\ell}=\chi^{\dagger}_{\ell}$ satisfy $\{\chi_{\ell},\chi_{\ell'}\}=2\delta_{\ell\ell'}$, $\{\chi_{\ell},c_{j,\ell'}\}=\{\chi_{\ell},c^{\dagger}_{j,\ell'}\}=0$, thus maintaining the correct anti-commutation relations \cite{Nuclear_Cramp__2013,Nuclear_Giuliano_2020}.

The open boundary condition (OBC) of the chains implies that $c_{\ell,0}=c_{\ell,\pm(N+1)}=0$. Consequently, in momentum space, we have the modes \cite{LIEB1961407}
\begin{equation}
    c_{\ell,j}=\sqrt{\frac{2}{L+1}}\sum_{k}\sin\left(|j|k\right)\psi_{\ell,k},
\end{equation}
where $k=k_n=\frac{n\pi}{L+1}$, with $n=1,2,\cdots,L$, and $L=N-1\approx N$ is the length of each chain (here we used the lattice constant $a= 1$ as length unit). Thus, in momentum space we find,
\begin{equation}
    H_0 = \sum_{k,\ell}\epsilon_{\ell,k}\psi^{\dagger}_{\ell,k}\psi_{\ell,k} + E_{\ell,0},
    \label{H0}
\end{equation}
where $\epsilon_{\ell,k}=h_{\ell}-\cos k$ is the energy dispersion relation of the fermionic excitations and $E_{\ell,0}=-h_{\ell}L/2$, which we can ignore since it just shifts the full spectrum by a constant. Notice that the Klein factors do not appear in the above equation and $[\chi_{\ell},H_0]=0$. Thus, they have no dynamics in the interaction picture.

Depending on the value of $h_{\ell}$, $H_{\ell}$ has a gapless ($h_{\ell}<1$) or a gapped spectrum ($h_{\ell}>1$). In the gapless case, the ground state $|\Upsilon_{\ell,0}\rangle$ of $H_{\ell}$ has a Fermi sea form \cite{coleman2015introduction}
\begin{equation}
    |\Upsilon_{\ell,0}\rangle = \prod_{k\leq k_{\ell,F}}\psi^{\dagger}_{\ell,k}|0\rangle,
\end{equation}
where $k_{\ell,F}=\arccos h_{\ell}$ is the Fermi momentum of the chain $\ell$ and $|0\rangle$ is the fermionic vacuum state. In the gapped case, the ground state is just the vacuum state $|\Upsilon_{\ell,0}\rangle = |0\rangle$, so that $\psi_{\ell,k}|\Upsilon_{\ell,0}\rangle=0$ for all $k$. As we will see in the following, this fact is crucial in determining the long-time decay of $W_{\te{ex}}(\tau)$. 

\begin{figure}
    \centering
    \includegraphics[scale=0.5]{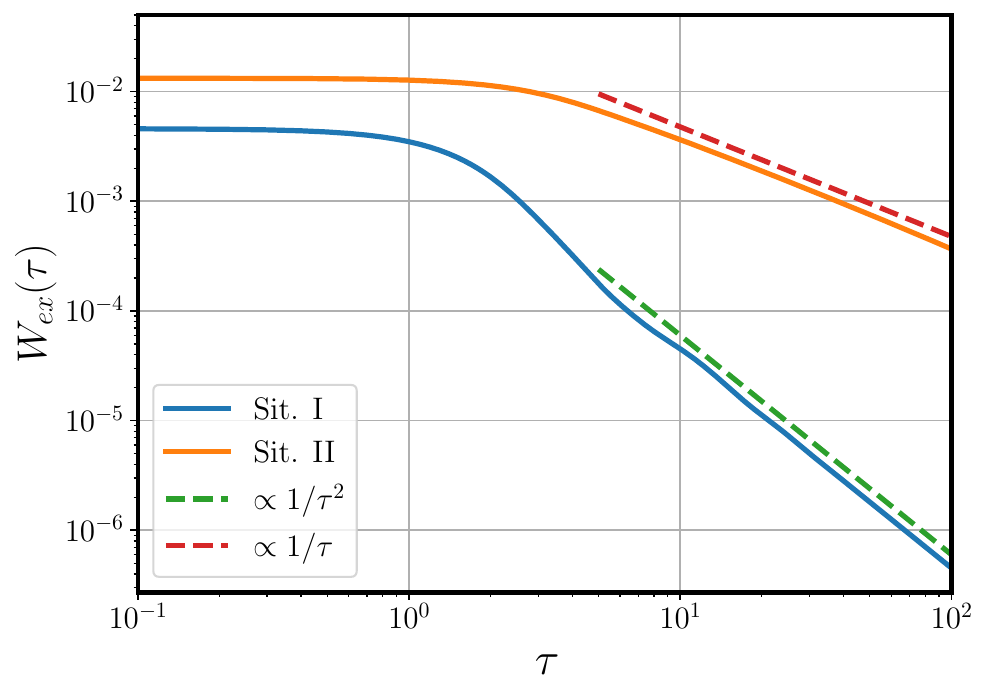}
    \caption{Excess work for $J_z=0$ ($XX$-case). The solid lines represent the numerical evaluation of the sums over $k,k'$ [Eqs.~\eqref{ex_work_sit_I} and \eqref{resp_fun_sit_II}] while the dashed ones the results obtained from the low-energy approach  [Eqs.~\eqref{ex_work_sit_I_result} and \eqref{ex_work_sit_II_result}]. Parameters are  $g_0=1/4$, $J=1$, $k_F=\pi/2$, $\Delta=1$ , $L=600$ and $\Lambda/v_F\sim 1$.}
    \label{fig_excess_work_XX}
\end{figure}

The coupling between the chains given in Eq. (\ref{coupling}) becomes in the fermionic representation
\begin{equation}
V=g_0\chi\sum_{k,k'}f_{kk'}\psi^{\dagger}_{1,k}\psi_{2,k'}+\te{H.c.},
\label{coupl_dif_chains}
\end{equation}
where $f_{kk'}=\frac{1}{L+1}\sin k\sin k'$ and $\chi=\chi_1\chi_2$. We see that $V$ promotes the tunneling of excitations  between the two chains. Furthermore, since $H_0$ represents a system of free fermions, see Eq. (\ref{H0}), it is easy to show that $e^{iH_0t}\psi_{\ell,k}e^{-iH_0t}=e^{-i\epsilon_{\ell,k}t}\psi_{\ell,k}$ and, thus, to obtain $V_I(t)$. 

In contrast to $H_0$, the coupling in Eq.~(\ref{coupl_dif_chains}) explicitly involves the Klein factors, though only through the operator $\chi$. As this operator commutes with the Hamiltonian (\ref{junction}) for all times, $[\chi,\mc H(\lambda)]=0$, we can choose to work in one of the two subspaces defined by the eigenvalues $\pm i$ of $\chi$ during the entire protocol \cite{Nuclear_Giuliano_2020}. Thus, hereafter, we set $\chi=i$ in Eq. (\ref{coupl_dif_chains}). Since the response function in Eq. (\ref{respon_fun}) involves a squared contribution in $V$, this choice does not have any measurable physical consequence, and hence we do not lose generality.

\paragraph*{Situation I: Gapless -- Gapped}

As a first example, we consider that one chain is in the gapless phase and the other one is in the gapped phase, $h_{1}<1$ and $h_{2}>1$.

As described in Sec.~\ref{sec_ex}, the first step to obtain the excess work is to evaluate the response function in Eq.~(\ref{respon_fun}). We find
\begin{equation}
    \Phi(t'-t)=2g^2_0\sum_{k\leq k_F,k'}f^2_{kk'}\sin\left[\epsilon_{kk'}(t-t')\right],
    \label{resp_fun_sit_I}
\end{equation}
where $\epsilon_{kk'}\equiv\epsilon_{1,k}-\epsilon_{2,k'}$ and $k_F$ is the Fermi momentum of chain $1$. The relaxation function $\Psi(t'-t)$, defined in Eq. (\ref{relax_fun}), can be directly obtained from the above result. Then, for the linear protocol, we obtain the excess work 
\begin{equation}
    W_{\te{ex}}(\tau)=2g^2_0\sum_{k\leq k_F,k'}\frac{f^2_{kk'}}{\epsilon^3_{kk'}}\left[\frac{\cos\left(\epsilon_{kk'}\tau\right)-1}{\tau^2}\right].
    \label{ex_work_sit_I}
\end{equation}
Since $\epsilon_{kk'}<0$, due to the restriction $k\leq k_F$, we see that $W_{\te{ex}}(\tau)\geq 0$, as expected \cite{Naze_2020}.

The double-$k$ summation in Eq. (\ref{ex_work_sit_I}) makes any analytical progress difficult. However, given that $\epsilon_{kk'}$ never vanishes due to the gap in the second chain, we expect that for long times $W_{\te{ex}}(\tau)$ scales as $1/\tau^2$, as predicted in Ref.~\cite{Soriani_2022}. Indeed, we will shortly see that this is confirmed by numerical calculations.

We now continue with a low-energy approach that captures the asymptotic long-time behavior of $W_{\te{ex}}(\tau)$ in Eq. (\ref{ex_work_sit_I}). In this limit, we can use that
\begin{eqnarray}
    \epsilon_{1,k}&\approx&v_F(k-k_F),\nonumber\\
    \epsilon_{2,k'}&\approx& \Delta,\nonumber\\
    f_{kk'}&\approx&\frac{\sin k_F}{L+1}k',
\end{eqnarray}
where $v_F=\sin k_F$ is the Fermi velocity and $\Delta\equiv h_2-1$ is the gap of the second chain. Substituting the above results into Eq. (\ref{ex_work_sit_I}) and taking the limit of semi-infinite chains $L\to \infty$, we  obtain
\begin{equation}
    W_{\te{ex}}(\tau\gg\Delta^{-1})\approx\frac{\Lambda^3}{3v^4_F}\left(\frac{g_0}{\pi}\right)^2\left[1-\frac{\Delta^2}{(\Lambda+\Delta)^2}\right]\frac{1}{(\Delta\tau)^2},
    \label{ex_work_sit_I_result}
\end{equation}
where  $\Lambda/v_F$ is a large momentum cutoff. As expected from our earlier analysis of Eq. (\ref{ex_work_sit_I}), $W_{\te{ex}}(\tau)$ scales as $1/\tau^2$ for long times. In Fig. \ref{fig_excess_work_XX} we show the numerical evaluation of Eq. (\ref{ex_work_sit_I}), which shows the $1/\tau^2$ scaling already for times $\tau \gtrsim 10J^{-1}$. As mentioned before, this corroborates the predictions of Ref.~\cite{Soriani_2022} but for a much more complex system that is partially gapless. Remarkably, the $1/\tau^2$ behavior of $W_{\te{ex}}(\tau)$ for larger $\tau$  shown in Eq. (\ref{ex_work_sit_I_result}) is protocol independent, see Appendix \ref{apend_B} for details.

\paragraph*{Situation II: Gapless -- Gapless}

As a second example, we consider that both chains are in the gapless phase and that, for simplicity, we have identical chains, i.e., $h_{1}=h_{2}\equiv h<1$. Similarly to above, we have the response function
\begin{equation}
    \Phi(t'-t)=4g^2_0\sum_{k\leq k_F,k'> k_F}f^2_{kk'}\sin[\epsilon_{kk'}(t-t')].
    \label{resp_fun_sit_II}
\end{equation}
Comparing the above result with Eq. (\ref{resp_fun_sit_I}), we notice an essential difference between them in the $k'$-summation interval. 

\begin{figure}
    \centering
    \includegraphics[scale=0.5]{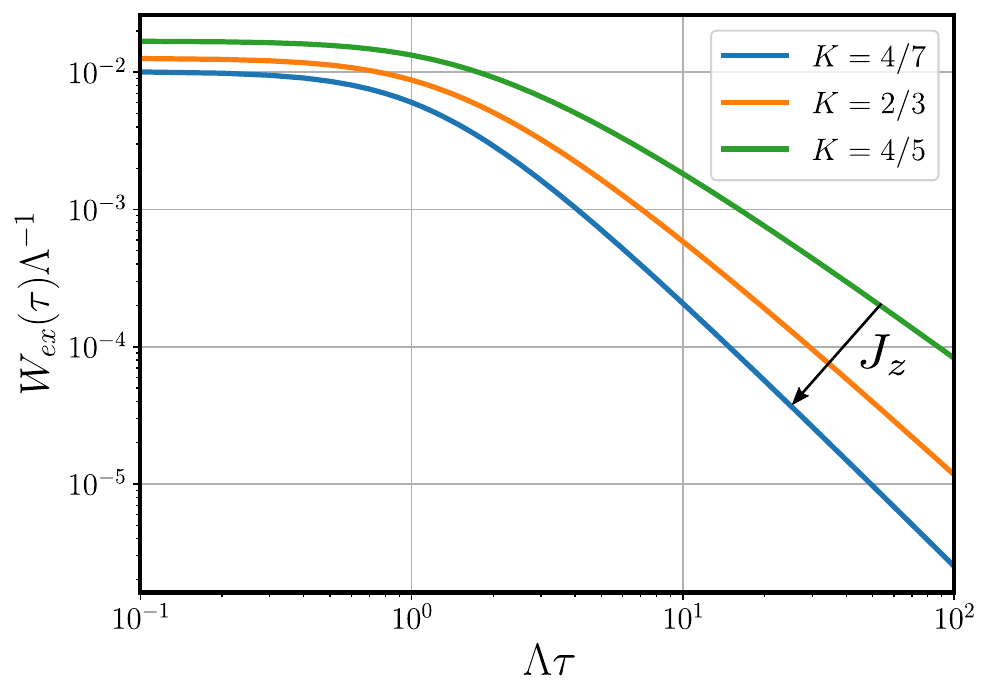}
    \caption{Excess work for $J_z\neq 0$ ($XXZ$-case) according to Eq. (\ref{ex_work_XXZ_result}). The Luttinger parameters $K=4/7$, $2/3$ and $4/5$ in the legend correspond to $J_z\approx0.92$, $1/\sqrt{2}$ and $0.38$, respectively. These values of $K$ were chosen so that all three cases in Eq. (\ref{ex_work_XXZ_result_2}) are illustrated. Parameters are   $g_0=1/4$ and $\Lambda\alpha\sim 1$.}
    \label{fig_excess_work_XXZ}
\end{figure}

In this case, since the large-chain limit implies that $\Delta k=k_{n+1}-k_n=\frac{\pi}{L+1}\to 0$, the energy difference $\epsilon_{kk'}$ vanishes close to $k_F$. Thus we expect a qualitatively different long-time behavior for $W_{\te{ex}}(\tau)$. To check this, we again apply a low-energy approach. Now we have $\epsilon_{kk'}\approx v_F(k-k')$ and $f_{kk'}\approx \sin^2 k_F/(L+1)$, which yields
\begin{equation}
    W_{\te{ex}}(\tau\gg J^{-1})\approx\frac{2}{\pi}\left(\frac{\tilde g_0}{v_F}\right)^2\frac{1}{\tau},
    \label{ex_work_sit_II_result}
\end{equation}
where $\tilde g_0= g_0\sin^2 k_F$. The above result demonstrates how the absence of a gap in the system drastically changes the long-time behavior of $W_{\te{ex}}(\tau)$. Interestingly, this result does not depend on the high-energy cutoff $\Lambda$ as Eq. (\ref{ex_work_sit_I_result}) does. In Fig.~\ref{fig_excess_work_XX} we show $W_{\te{ex}}(\tau)$, both from the exact numerical calculation of Eq.~\eqref{resp_fun_sit_II} and the asymptotic result in Eq.~\eqref{ex_work_sit_II_result}. We observe that the $1/\tau$ scaling is exhibited already for times $\tau \gtrsim 10J^{-1}$. This result does not contradict what was shown in Ref.~\cite{Soriani_2022} since here the system is completely gapless.

\begin{figure*}
    \centering
    \includegraphics[scale=0.35]{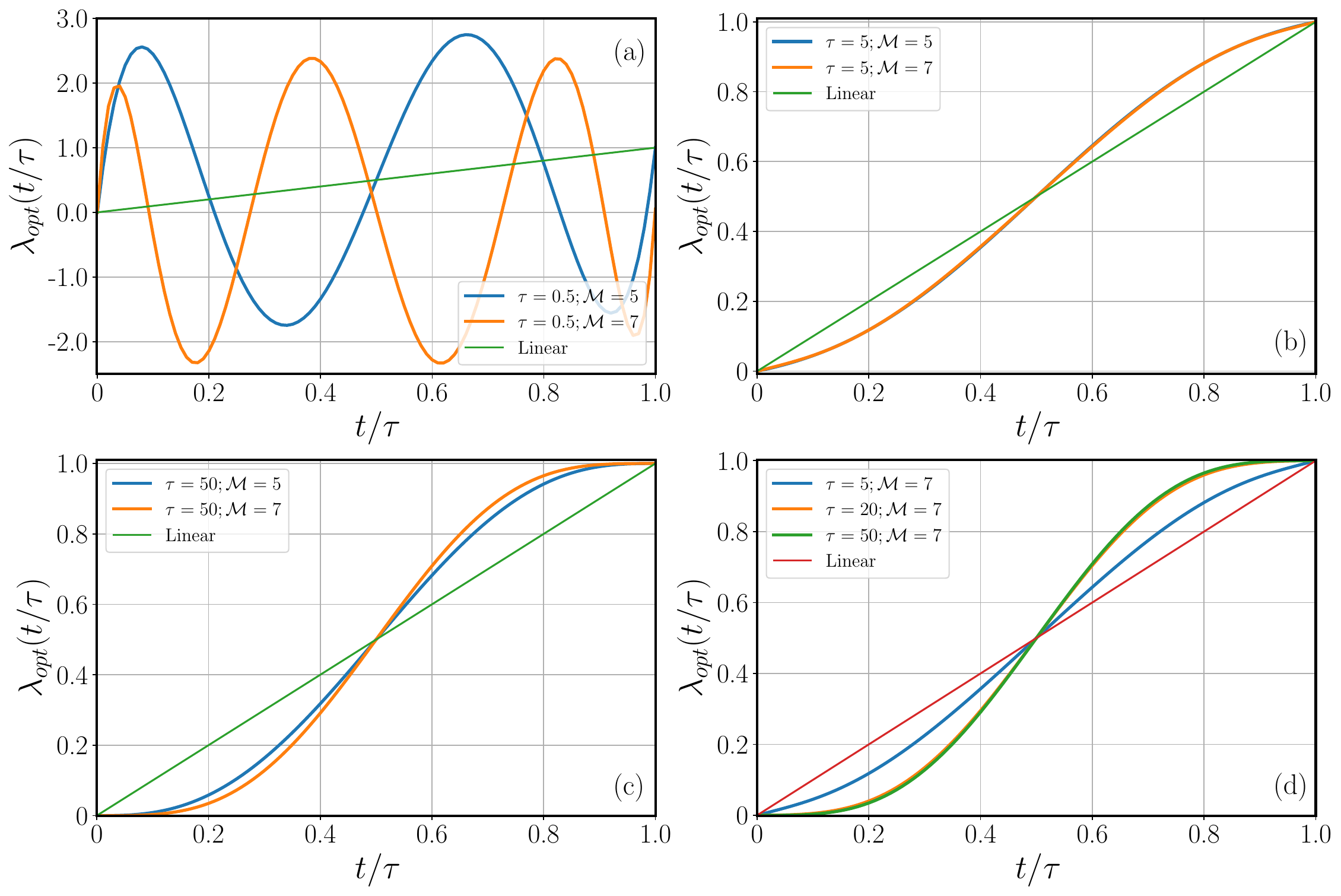}
    \caption{Optimal protocols for situation I (\textit{gapless -- gapped}). In panels (a-c) we show the optimal protocols for $\tau=0.5$, $5$ and $50$ (in units of $J^{-1}$), respectively, for $\mc M=5$ and $7$. Panel (d) shows the evolution of $\lambda_{\te{opt}}(t)$ when we increase $\tau$ while keeping $\mc M=7$. The optimal protocol for $\tau=0.5$ and $\mc M=7$ is multiplied by $0.05$ for clarity. For comparison, the linear protocol is plotted in all panels. Parameters are  $g_0=1/4$, $J=1$, $k_F=\pi/2$, $\Delta=1$ and $L=100$.}
    \label{fig_opt_sit_I}
\end{figure*}

In Sec.\ref{sec_opt_prot}, based on a field theory description, we will see that the $1/\tau$ scaling of $W_{\te{ex}}(\tau \gg J^{-1})$ shown in Eq. (\ref{ex_work_sit_II_result}) is also protocol independent.

\subsection{Case 2: $J_z\neq0$ -- the $XXZ$ model}
\label{sec_XXZ}

As a fundamentally different scenario, we now analyze chains with a $J_z\neq0$ anisotropy. To this end, we assume that both chains are in the antiferromagnetic critical phase, $0<J_z<1$, and we set $h_{\ell}=0$ \cite{Haldane,Sirker_2006}. This assumption does not affect the generality of our results, since a finite $h_{\ell}$, as long as we stay in the critical phase, just shifts the value of $k_F$. It follows then that now $H_0$ has spin inversion symmetry $S^a_{j}\to-S^a_{j}$, which translates into particle-hole symmetry of the fermionic representation $c_{\ell,j}\to(-1)^j c^{\dagger}_{\ell,j}$, and, consequently, implies that  $k_{1,F}=k_{2,F}=\pi/2$. 

In this case, $J_z$ introduces a quartic interaction term between the fermions, see the last of Eqs.~(\ref{J_W_trans}), and the chain Hamiltonians of Eq.~\eqref{H1} are no longer easily diagonalizable \cite{giamarchi2004quantum}. We therefore must apply a more powerful field-theoretical bosonization technique \cite{giamarchi2004quantum,gogolin2004bosonization}. 

\paragraph*{Field theory approach}

In the fermionic representation, $H_0$ assumes the following form 
\begin{eqnarray}
    H_0=&-&\frac{1}{2}\sum_{\ell,j}\left(c^{\dagger}_{\ell,j+1}c_{\ell,j}+\te{H.c.} \right)\nonumber\\
    &&+J_z\sum_{\ell,j}(n_{\ell,j}-1/2)(n_{\ell,j+1}-1/2),
\end{eqnarray}
where $n_{\ell,j}=c^{\dagger}_{\ell,j}c_{\ell,j}$ is the number operator at site $j$ of chain $\ell$. Hereafter we work in the limit of semi-infinite chains $L\to \infty$. 

Focusing on the low-energy sector of the chains, we take the continuum limit and expand the fermionic field operators in terms of right (R) and left (L) movers as \cite{giamarchi2004quantum}
\begin{equation}
c_{\ell,j}\sim c_{\ell}(x)=e^{ik_{F}x}\psi_{R,\ell}(x) + e^{-ik_{F}x}\psi_{L,\ell}(x),
\label{L_R_expan}
\end{equation}
where $c_{1}(x)$ is defined for $x<0$  and $c_{2}(x)$  for $x>0$. The OBC on the chains, $c_{\ell}(0)=0$, can be cast in the form \cite{Fabrizio}
\begin{equation}
    \psi_{L,\ell}(x)=-\psi_{R,\ell}(-x).
    \label{OBC_constr}
\end{equation}
These constraints allow us to work with a single chiral fermionic field operator for each chain redefined in the domain $x\in \mathbbm R$ \cite{Fabrizio}. Thus, it is convenient to work with $\psi^\dagger_{1}(x)\equiv\psi^\dagger_{L,1}(-x)$ and $\psi^\dagger_{2}(x)\equiv\psi^\dagger_{R,2}(x)$ \cite{Helena}.

The low-energy modes of $H_0$ are described by the Hamiltonian \cite{Fabrizio}
\begin{eqnarray}
H_{0}&=&\sum_{\ell}\int dx\,\Big\{v_F\psi^\dagger_{\ell}(x)(-i\partial_x)\psi^{\phantom\dagger}_{\ell}(x) \nonumber\\
&&+\: J_z \left[\rho^2_{\ell}(x)+2\rho_{\ell}(x)\rho_{\ell}(-x)\right]\Big\}+H_{\te{Umk},\ell}\nonumber,\\ 
\label{HJ}
\end{eqnarray}
where $\rho_{\ell}(x)=:\psi^\dagger_{\ell}(x)\psi^{\phantom\dagger}_{\ell}(x):$ is a density operator ($:\cdots:$ denotes normal ordering) and $H_{\te{Umk},\ell}$ describes the Umklapp terms, known to open a gap in the system when $|J_z|>1$ \cite{giamarchi2004quantum}. Since we are in the critical phase of the $XXZ$ chain ($0<J_z<1$), this term consists of an irrelevant perturbation at low energies (in the renormalization group sense) and thus can be neglected \cite{giamarchi2004quantum,gogolin2004bosonization}. 

\begin{figure*}
    \centering
    \includegraphics[scale=0.35]{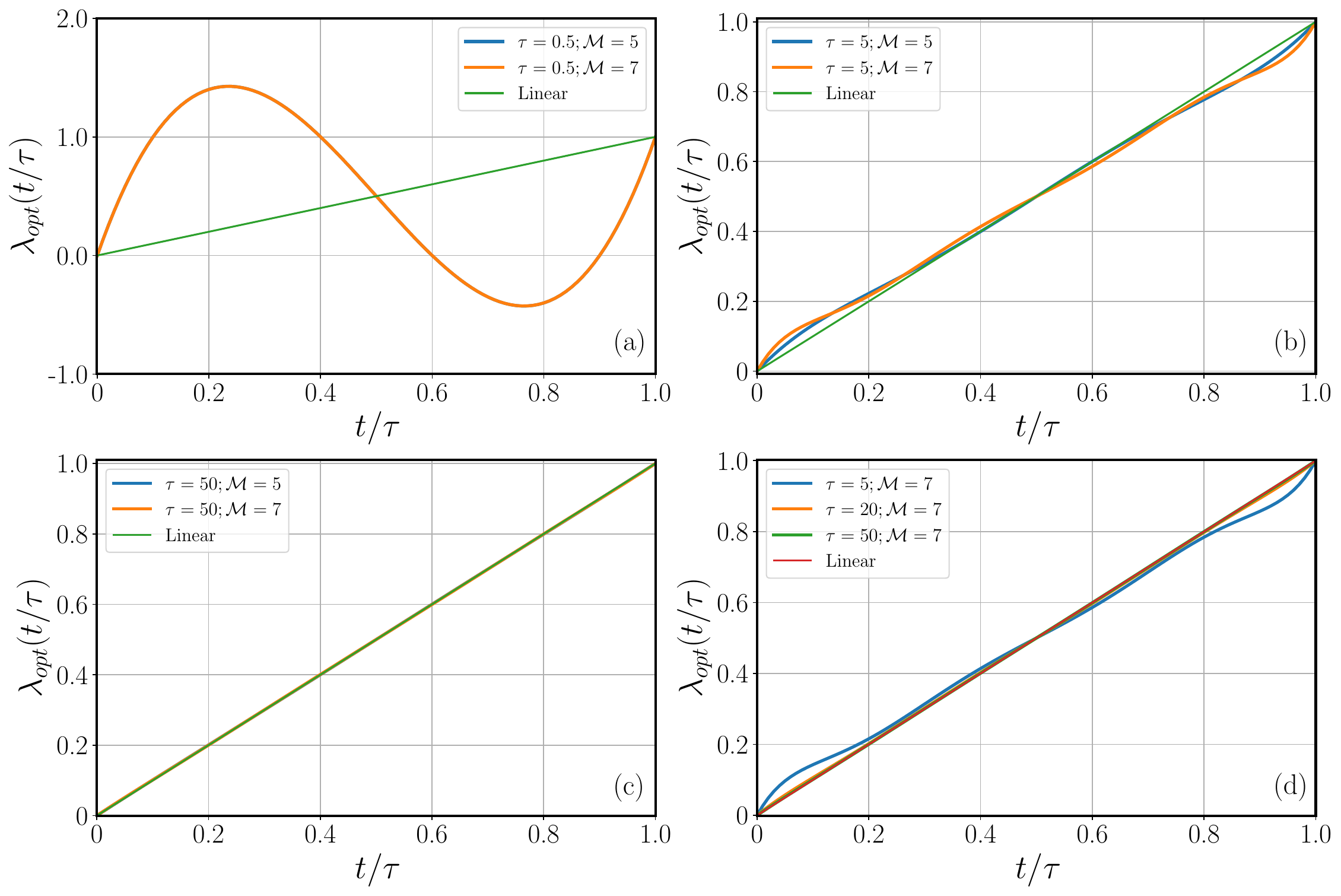}
    \caption{Optimal protocols for the situation II (\textit{gapless -- gapless}). In panels (a-c) we show the optimal protocols for $\tau=0.5$, $5$ and $50$ (in units of $J^{-1}$), respectively, for $\mc M=5$ and $7$. Panel (d) shows the evolution of $\lambda_{\te{opt}}(t)$ when we increase $\tau$ while keeping $\mc M=7$. For comparison, the linear protocol is also plotted in all panels. Parameters are  $g_0=1/4$, $J=1$, $k_F=\pi/2$ and $L=100$.}
    \label{fig_opt_sit_II}
\end{figure*}

In Eq. (\ref{HJ}) we are left with a Luttinger liquid (LL) model, with OBC, for each $XXZ$ chain \cite{giamarchi2004quantum}. This model can be solved by applying Abelian bosonization \cite{Fabrizio,giamarchi2004quantum,gogolin2004bosonization}. In this approach, the fermionic field operators and the density operators are written in terms of auxiliary bosonic fields $\phi_{\ell}(x)$ as \cite{Fabrizio,gogolin2004bosonization}
\begin{eqnarray}
\psi_{\ell}(x)&\sim&\frac{1}{\sqrt{2\pi\alpha}}F_{\ell}e^{-i\sqrt{\pi}\phi_{\ell}(x)}\label{field_boso},  \\
\rho_{\ell}(x)&\approx&-\frac{1}{\sqrt{4\pi}}\partial_x\phi_{\ell}(x),
\label{densi_boso}
\end{eqnarray}
where  $\alpha\sim v_F/\Lambda$ is a short-distance cutoff and $F_{\ell}$ are different Klein factors  that ensure the anticommutation relations between the fermions. In terms of its normal modes $\eta_{\ell,q}$, where $q>0$ are momenta, $\phi_{\ell}(x)$  has the form \cite{Fabrizio}
\begin{equation}
\phi_{\ell}(x)=\frac{1}{\sqrt{L}}\sum_{q>0}\frac{e^{-\frac{\alpha }{2}q}}{\sqrt{q}}[ z_{q}(x)\eta_{\ell,q} + z^{*}_{q}(x)\eta^{\dagger}_{\ell,q}],    \label{modeexp}
\end{equation}
where $z_{q}(x) = (1/\sqrt{K}) \cos(qx) + i\sqrt{K}\sin(qx)$, with $K$ the Luttinger parameter \cite{giamarchi2004quantum}.

Then, the interacting Hamiltonian in Eq.~\eqref{HJ} can be expressed as a \emph{quadratic} diagonal form in terms of the normal modes as 
\begin{equation}
    H_{0}=u\sum_{\ell,q>0}q\eta^{\dagger}_{\ell, q}\eta^{\phantom\dagger}_{\ell, q}, \label{Hbos}
\end{equation} 
where $u$ is the velocity of the bosonic modes. The important quantity in the bosonized theory is the Luttinger parameter $K$. For small $J_z$, $K\approx 1 - \frac{2}{\pi}J_z$, so $K=1$ in the non-interacting $XX$-case. The exact expression of $K$ for any value of $J_z$ is known from the Bethe ansatz solution \cite{Sirker_2006}, and is given by $K=\frac{\pi/2}{\pi-\arccos J_z}$. We see that $1/2 < K \leq 1$ in the critical region we are interested in $0\leq J_z<1$ \cite{giamarchi2004quantum}. Furthermore, the velocity $u$ is also known from the Bethe Ansatz solution, $u=\frac{\pi/2\sqrt{1-J^2_z}}{\arccos J_z}$ \cite{Sirker_2006}. Finally, within the low-energy theory, the ground state $\left|\Upsilon_{\ell,0}\right\rangle$ of  the $XXZ$ chain corresponds to the bosonic vacuum, $\eta_{\ell,q}\left|\Upsilon_{\ell,0}\right\rangle=0$ for all $q>0$ \cite{giamarchi2004quantum}. 

For the coupling $V$ between the chains, given in Eq.~(\ref{coupling}), we need only the field operators at the sites $j=\pm 1$. Using the low-energy expansion (\ref{L_R_expan}) and the OBC constraint (\ref{OBC_constr}), we obtain that $c_{2,1}\sim2i\sin k_F\,\psi_{2}(0)$ and $c_{1,-1}\sim 2i\sin k_F\,\psi_{1}(0)$ \cite{Helena}, remembering that $k_F=\pi/2$. Then, using Eq. (\ref{field_boso}), $V$ has the following bosonized form
\begin{equation}
    V\sim\bar g_0\cos\left[\sqrt{2\pi}\phi^{-}(0)\right],
    \label{coup_boso}
\end{equation}
where $\bar g_0=2g_0/(\pi\alpha)$ and we defined the odd/even combinations $\phi^{\pm}=(\phi_{2}\pm\phi_{1})/\sqrt{2}$. In the above equation, we omitted the Klein factors $F_{\ell}$. 

As pointed out before, a term like $S^{z}_{-1}S^{z}_{1}$ could also be considered in Eq.~(\ref{coup_boso}). In the bozonization representation, such a term [see Eq.~(\ref{densi_boso})] has the form $\propto\partial_x\phi_{1}(0)\partial_x\phi_{2}(0)$, whose scaling dimension is $d^{\prime}=2$. The operator in Eq. (\ref{coup_boso}), in contrast, has scaling dimension $d=1/K$ which, in our range of interest of $J_z$, is always smaller than $2$. Since terms with the lowest scaling dimension are dominant, the leading contribution comes from Eq. (\ref{coup_boso}) \cite{Furusaki_1}. 

Since the scaling dimension of $V$ is a nonuniversal number, we expect that it produces an anomalous long-time power-law for $W_{\te{ex}}(\tau)$, in contrast to the $1/\tau$ power-law obtained in the $XX$-case (see Fig.~\ref{fig_excess_work_XX}). 

\paragraph*{Excess work}

The low-energy approach described above allows us to also obtain a closed expression for the relaxation function $\Psi(s)$. After evaluating the two-point correlation function $\langle V_I(t')V_I(t)\rangle_0$ in the bosonized theory with standard methods \cite{boso_correl}, we obtain 
\begin{widetext}
\begin{equation}
    \Psi(s)=\frac{\Lambda}{(2d-1)}\left(\frac{2g_0}{\pi\Lambda\alpha}\right)^2\frac{\cos(2d\arctan\Lambda s)+\Lambda s\sin(2d\arctan\Lambda s)}{(1+\Lambda^2s^2)^{d}},
    \label{relax_func_XXZ}
\end{equation}
\end{widetext}
where, again, $d=1/K$ is the scaling dimension of $V$ in Eq. (\ref{coup_boso}). The result of Eq.~(\ref{relax_func_XXZ}) shows that $\Psi(s)$ scales as $1/s^{2d-1}$ for long times, which tells us that the anisotropy introduced by $J_z$, correspondingly the interactions between the fermions, cause the system to relax much faster to the equilibrium state ($d=1$ when $J_z=0$). As we will see below, this fact is reflected in the $W_{\te{ex}}(\tau)$ behavior.

\begin{figure*}
    \centering
    \includegraphics[scale=0.35]{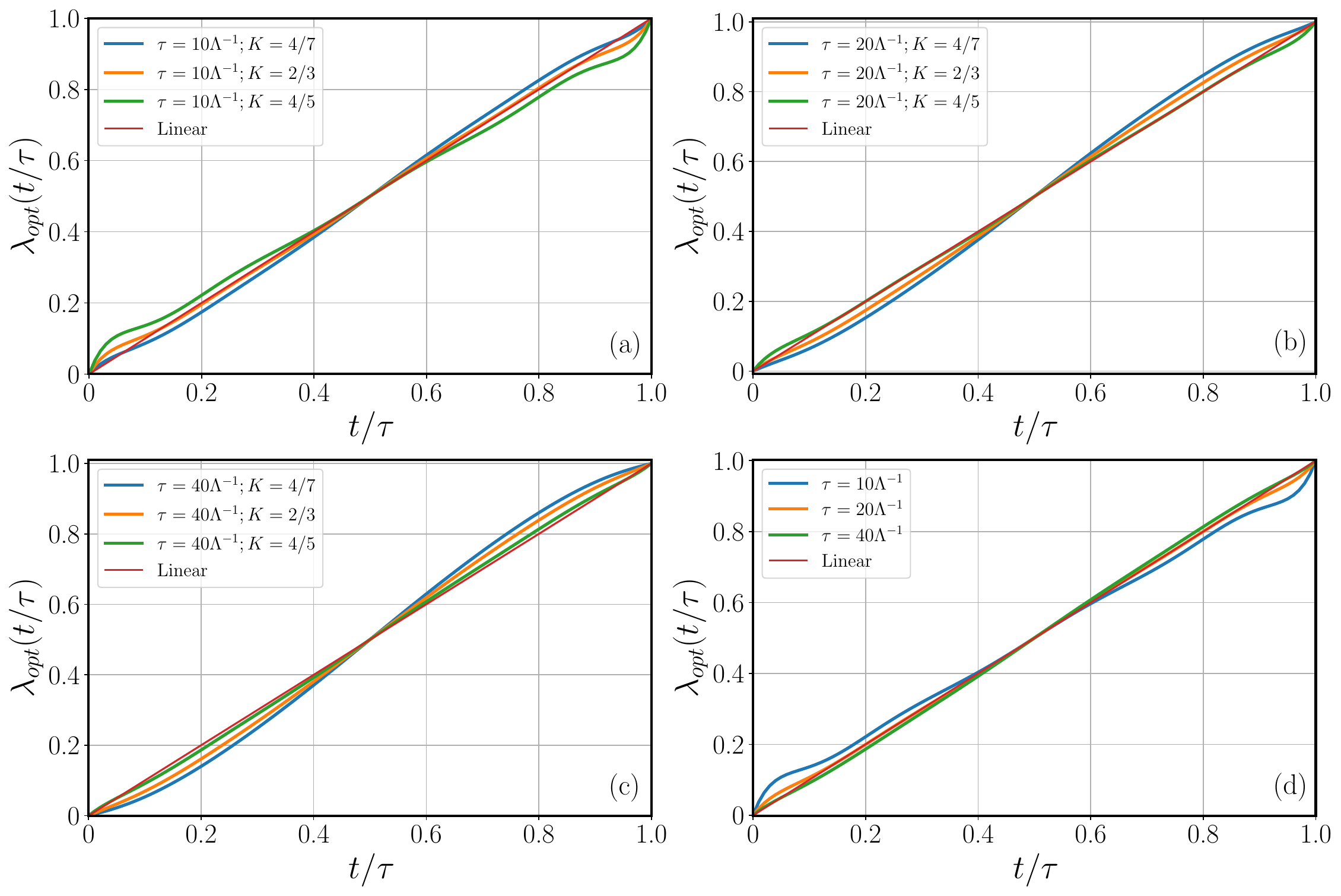}
    \caption{Optimal protocols for the $XXZ$-case. In panels (a-c) we show the optimal protocols for $\tau=10\Lambda^{-1}$, $20\Lambda^{-1}$ and $40\Lambda^{-1}$, respectively, for different values of $K$. Panel (d) shows the evolution of $\lambda_{\te{opt}}(t)$ when we increase $\tau$ while keeping $K=4/5$. Once again, for comparison, we show the linear protocol in all panels. The Luttinger parameters $K=4/7$, $2/3$ and $4/5$ correspond to $J_z\approx0.92$, $1/\sqrt{2}$ and $0.38$, respectively. These values of $K$ were chosen so that all three regimes in Eq.~(\ref{ex_work_XXZ_result_2}) are illustrated. Parameters are  $g_0=1/4$, $k_F=\pi/2$ and $\mc M=9$.}
    \label{fig_opt_XXZ}
\end{figure*}

After again deriving the relaxation function, the excess work in Eq.~(\ref{excesswork}), for the linear protocol, can be immediately obtained. It is given by
\begin{widetext}
\begin{eqnarray}
    W_{\te{ex}}(\tau)/\Lambda&=&f(d)\left(\frac{2g_0}{\pi\Lambda\alpha}\right)^2\frac{(1+\Lambda^2\tau^2)^{d}+(-1+3\Lambda^2\tau^2)\cos(2d\arctan\Lambda\tau)+\Lambda\tau(-3+\Lambda^2\tau^2)\sin(2d\arctan\Lambda\tau)}{(\Lambda\tau)^2(1+\Lambda^2\tau^2)^{d}},\nonumber\\
    \label{ex_work_XXZ_result}
\end{eqnarray}
where $f(d)\equiv1/\left[(2d-1)(d-1)(2d-3)\right]$. This result is shown in Fig. \ref{fig_excess_work_XXZ} for different values of $K$. As anticipated, the anisotropy introduced by $J_z$ induces a much richer behavior for $W_{\te{ex}}(\tau)$, such that the simple $1/\tau$ scaling of the $XX$-case is lost. Indeed, for long times we  now have
\begin{eqnarray}
    W_{\te{ex}}(\tau\gg \Lambda^{-1})/\Lambda&\approx&\begin{cases}
       f(d)\left(\frac{2g_0}{\pi\Lambda\alpha}\right)^2\frac{1}{(\Lambda\tau)^2}, &\quad 1/2<K<2/3\\
       \left(\frac{2g_0}{\pi\Lambda\alpha}\right)^2\frac{\ln[\Lambda\tau]}{(\Lambda\tau)^2}, &\quad K=2/3 \\
      
      f(d)\left(\frac{2g_0}{\pi\Lambda\alpha}\right)^2\frac{\sin(K^{-1}\pi)}{(\Lambda\tau)^{\frac{2}{K}-1}}, &\quad 2/3<K\leq1.\\
     \end{cases}
     \label{ex_work_XXZ_result_2}
\end{eqnarray}
\end{widetext}
We see that the more correlated the system is, that is, the larger $J_z$, the faster $W_{\te{ex}}(\tau)$ scales to zero, as a consequence of a much faster relaxation of $\Psi(\tau)$ towards equilibrium. Also, in the range $2/3 < K <1$, $W_{\te{ex}}(\tau)$ scales with a non-universal power $\nu=2/K -1$, determined by the anomalous scaling dimension of the coupling in Eq.~(\ref{coup_boso}) when $J_z\neq 0$. For $1/2 < K \le 2/3$, the power ``sticks'' to $\nu=2$. In Ref.~\cite{NFL_dynamics}, the authors obtained a similar result, that is, the excitation energy, defined as $\Delta E(\tau)=\langle\Upsilon(\tau)|\mc H(\tau)|\Upsilon(\tau)\rangle-\langle \mc H(\tau)\rangle_0\sim \tau^{-\eta}$,  scaling with a noninteger power $\eta$ after a linear protocol, but in the context of the dynamics of a non-Fermi liquid system.

Finally, we highlight that the result in Eq. (\ref{ex_work_XXZ_result_2}) clearly shows how complex gapless many-body systems can lead to non-trivial behavior for $W_{\te{ex}}(\tau)$. In our particular case, this even suggests a recipe for achieving quasi-adiabaticity in finite time: tayloring  ever more strongly interacting systems in order to obtain a faster decay of the excess work with the process duration. 

\section{Optimal protocols}
\label{sec_opt_prot}

After discussing the behavior of $W_{\te{ex}}(\tau)$ for a specific protocol, we now turn to solving the actual optimal control problem. Specifically, we determine the optimal protocol $\lambda_{\te{opt}}(t)$ for the junctions described by Eq. (\ref{junction}), i.e., the protocol which leads to the least amount of excess work in finite time $\tau$. 

To this end, we follow the strategy put forth by some of us in Ref.~\cite{Marcus_2}. In this approach we expand the time derivative of the protocol $\lambda(t)$ in terms of the Chebyshev polynomials $T_l(x)$ \cite{Chebyshev},
\begin{equation}
    \dot \lambda(t)=\sum_{l=1}^{\mc M}a_lY^{l}_{\mc M}T_l\left(2t/\tau - 1\right),
    \label{prot_expansion}
\end{equation}
and determine the coefficients $a_l$ that minimize the excess work of Eq.~(\ref{excesswork}), subject to the boundary conditions $\lambda(0)=0$ and $\lambda(\tau)=1$. These conditions are implemented through Lagrange multipliers \cite{Marcus_2}. The factors $Y^{l}_{\mc M}$ \cite{regularize_exp} in Eq.~(\ref{prot_expansion}) are needed to regularize the expansion, since we consider only the first $\mc M$ modes \cite{Chebyshev}.

We start by considering the $XX$-case, $J_z=0$. After solving the minimization problem, we obtain the protocols shown in Fig.~\ref{fig_opt_sit_I} for situation I, \textit{gapless -- gapped}. We find that the shape of the optimal protocol is strongly dependent on the time duration for $\tau<1$ (in units of $J^{-1}$), see Fig.~\ref{fig_opt_sit_I} (a), and exhibits an oscillatory behavior. When $\tau$ increases, a smooth behavior is obtained, see Fig.~\ref{fig_opt_sit_I} (b-d), and, for sufficiently long times $\tau \gtrsim 20$, the derivative of $\lambda_{\te{opt}}(t)$ approaches zero at the initial and final points, in agreement with the optimal protocols obtained from Adiabatic Perturbation Theory, as discussed in Ref.~\cite{Soriani_2022}. There, the same kind of optimal protocol was obtained for the quantum Ising chain (QIC) deep in the paramagnetic phase, which is also a gapped system. We also see that, for $\tau \gtrsim 5$, the expansion in Eq.~(\ref{prot_expansion}) converges just with a few Chebyshev polynomials.

When both chains are in the $XX$ gapless phase, the optimal protocols have the shapes shown in Fig.~\ref{fig_opt_sit_II}. As in the previous situation, the corresponding optimal protocol for short duration has an oscillatory behavior, but now with a smaller frequency, see Fig.~\ref{fig_opt_sit_II} (a). When $\tau$ is increased, the amplitude of this oscillation is rapidly suppressed and for $\tau \gtrsim 5$ (in units of $J^{-1}$) it is only seen close to the final and initial points, see Fig.~\ref{fig_opt_sit_II} (b).

Interestingly, if we continue to increase $\tau$ above $20$, the optimal protocol approaches the linear protocol, see Fig.~\ref{fig_opt_sit_II} (c-d). This can be understood from the low-energy approach employed in Sec.~\ref{sec_XX} or, equivalently, taking the noninteracting limit, $d=1$,
of Eq.~(\ref{relax_func_XXZ}). Following the latter approach, we find
\begin{eqnarray}
    \Psi(s)&=&v_F\left(\frac{2g_0}{\pi v_F}\right)^2\frac{\alpha}{\alpha^2+v^2_Fs^2},\nonumber\\
    &\overset{\alpha\to 0}{=}&\pi\left(\frac{2g_0}{\pi v_F}\right)^2\delta(s),
    \label{relax_fun_long_times}
\end{eqnarray}
where the second equality must be understood as being valid strictly for very long times $s$. Using the above result in Eq.~(\ref{excesswork}),
\begin{eqnarray}
    W_{\te{ex}}(\tau\gg\Lambda^{-1})&=&\frac{2}{\pi}\left(\frac{g_0}{v_F}\right)^2\frac{1}{\tau}\int_0^1 ds[\dot \lambda(s)]^2.
    \label{opt_exc_sit_I}
\end{eqnarray}
It is apparent that the protocol $\lambda(s)$ which satisfies the boundary conditions and minimizes Eq.~\eqref{opt_exc_sit_I} is the linear protocol \cite{Opt_LL}. Furthermore, Eq. (\ref{opt_exc_sit_I}) shows clearly that, for situation II, $W_{\te{ex}}(\tau\gg\Lambda^{-1})$ scales as $1/\tau$ for \textit{any} protocol. This result is a consequence of the marginality of the operator $V$ in the noninteracting case. 

We now turn to the case with a $J_z$-anisotropy. For this case, the optimal protocols are shown in Fig.~\ref{fig_opt_XXZ}. We observe that the presence of correlations in the chains changes the optimal protocol, so that the linear protocol is no longer optimal even for very long times $\tau$. However, for a small $J_z$, which corresponds to $K\lesssim 1$, the optimal protocol is fairly close to the linear one, differing from it only near the initial and final points. Increasing $J_z$ continuously, the optimal protocol becomes increasingly more distant from the linear protocol. In Ref.~\cite{Opt_LL} the authors obtained optimal protocols very similar to our findings but in the context of achieving the interacting ground-state of a LL starting with a non-interacting LL. 

In comparison with the linear protocol, situation I, \textit{gapless--gapped}, has the most distant optimal protocol, see Fig.~\ref{fig_opt_sit_I}. Furthermore, besides having the same $1/\tau^2$ power-law decay for the excess work, the optimal protocol for the QIC deep in the paramagnetic phase, obtained in reference \cite{Soriani_2022}, is the same as the one we found here in our case and shown in Fig.~\ref{fig_opt_sit_I} (c) (see Fig. 3 (a) of \cite{Soriani_2022}). In order to clarify this, let us look at the relaxation function of each system. Applying a low-energy approximation, one finds for long times $s$
\begin{eqnarray}
    \Psi^{\te{QIC}}(s\gg \tau_R)&\sim&\frac{\cos \Gamma s}{s^{3/2}},
    \label{relax_func_Ising} \\
    \Psi(s\gg \tau_R)&\sim&\frac{\cos \Delta s}{s^{5/2}},
    \label{relax_func_sit_I_long_times}
\end{eqnarray}
where $\Gamma$ is the gap deep in the paramagnetic phase (see Eq. A7 of \cite{Soriani_2022}). The extra power $s^{-1}$ in our case, i.e., a faster relaxation, is a consequence of the boundary coupling between the two chains. Since these two functions have different power-law decays, it suggests that the exponents controlling their decays are not the determining factor in the behavior of $W_{\te{ex}}(\tau\gg 1)$, as a naive analysis of Eq.~(\ref{excesswork}) might suggest. Indeed, looking at the spectral expression of $W_{\te{ex}}(\tau)$ in Eq.~(\ref{ex_work_sit_I}), which has a general structure (system-dependent details are encoded in $f_{kk'}$ and $\epsilon_{kk'}$), we can see that, as long as $\epsilon_{kk'}\neq0$, i.e., for a gapped system, the $1/\tau^2$ power law holds. This explains why the two systems show the same behavior, $W_{\te{ex}}(\tau\gg 1)\propto1/\tau^2$. 

Regarding the optimal protocol $\lambda_{\te{opt}}(t)$, we notice the crucial importance of the oscillatory behaviors in Eq. (\ref{relax_func_Ising}) and Eq. (\ref{relax_func_sit_I_long_times}) and thus of a finite gap in the system as follows. When both chains are in the critical LL phase, with, e.g., $K=4/5$ or $K=4/7$, the relaxation function in Eq.~(\ref{relax_func_XXZ}) has the long-time behaviors $\Psi(s\gg\Lambda^{-1})\propto s^{-3/2}$ and $\Psi(s\gg\Lambda^{-1})\propto s^{-5/2}$, respectively. However, the corresponding optimal protocols are clearly different [compare Figs.~\ref{fig_opt_XXZ} and \ref{fig_opt_sit_I} (c)] even for very long times. Therefore, we conclude that the presence of an energy gap in the system also plays an important role in determining $\lambda_{\te{opt}}(t)$.

\section{Concluding remarks}
\label{sec_discussion}

In this analysis, we have studied the excess work $W_{\te{ex}}(\tau)$ for establishing a weak junction between two quantum spin-$1/2$ chains in finite time.
First, we considered that the junction coupling strength increases linearly in time, i.e., $\lambda(t)=t/\tau$. For very short times $\tau$, we found that the excess work scales in time as $\propto(1-\text{const.}\tau^2)$, where the system details are encoded only in the constant pre-factor. Thus, this $\tau^2$ behavior is independent of which phase the chains are in. On the other hand, for long times $\tau$, the excess work depends on the chain phases. For the $XX$-case, i.e., $J_z=0$, $W_{\te{ex}}(\tau\gg 1)$ scales as $1/\tau$ when both chains are in the $XX$ gapless phase and as $1/\tau^2$ when one of the chains is in the $XX$ gapless phase and the other one in the $XX$ gapped phase. Notably, both of these behaviors are \textit{protocol independent} [see Eq.~(\ref{opt_exc_sit_I}) and Appendix \ref{apend_B}]. 

The $1/\tau^2$ power-law scaling for gapped systems was also recently obtained, by some of us, when we analyzed general (not necessarily weak), but slow (quasi-adiabatic) protocols in the transverse field of the QIC \cite{Soriani_2022}. 
Our results suggest that a finite spectral gap plays a key role in the $W_{\te{ex}}(\tau)$ behavior, as anticipated in Ref.~\cite{Soriani_2022}. Moreover, it shows that even when the system of interest is partially gapless, the scaling predicted by Ref.~\cite{Soriani_2022} applies.

On the other hand, in the presence of $J_z$-anisotropy, $W_{\te{ex}}(\tau\gg 1)$ has a richer scaling behavior; see Eq.~(\ref{ex_work_XXZ_result_2}). In particular, when $2/3< K<1$, corresponding to  $(\sqrt{2})^{-1}<J_z<1$, the excess work scales as a nonuniversal power law $1/\tau^{\nu}$, where 
\begin{equation}
    \nu=2/K-1.
\end{equation}
When compared with the simple $1/\tau$ power-law of the $XX$-case, this result highlights how complex gapless many-body systems show non-trivial behavior for $W_{\te{ex}}(\tau)$.
The energy of the system after the protocol, more precisely the excitation energy $\Delta E(\tau)=\langle\Upsilon(\tau)|\mc H(\tau)|\Upsilon(\tau)\rangle-\langle \mc H(\tau)\rangle_0$, was also shown to exhibit time scaling as $\sim \tau^{-\eta}$ with a noninteger exponent $\eta$ in a non-Fermi liquid system \cite{NFL_dynamics}.

Moreover, we also saw that $W_{\te{ex}}(\tau)$ decreases when we increase $J_z$ for fixed duration $\tau$ (see Fig.~\ref{fig_excess_work_XXZ}), which tells us that increasing the interactions in the system favors a quasi-adiabatic evolution. It is important to mention that non-universal quantities in the field-theory approach, such as the cutoff $\Lambda$, can explicitly depend on the interaction also, as was recently seen in \cite{Moa_2}. In any case, the results shown in Fig.~\ref{fig_excess_work_XXZ} illustrate how strongly correlated systems can play an important role in the search for shortcuts to adiabaticity and quantum control. 

Concerning the optimal protocol to establish the junction, we saw that it also depends on which phases the chains are in. When both chains are in the $XX$ gapless phase, the optimal protocol is the linear protocol for $\tau \gtrsim 20$ (in units of $J^{-1}$). Thus, Eq.~(\ref{ex_work_sit_II_result}) represents the minimal amount of energy that must be spent to establish the junction in finite long duration. When one of the chains is in the gapped phase, we found the same optimal protocol that was obtained for the gapped system considered in Ref.~\cite{Soriani_2022}. Our analyses suggest that this optimal protocol is general for gapped systems. In the presence of a finite but weak $J_z$-anisotropy, $\lambda_{\te{opt}}(t)$ differs slightly from the linear protocol near the initial and final points, see Fig.~\ref{fig_opt_XXZ} (d).

Finally, as future prospects, it would be interesting to extend our analysis to other types of quantum spin chains and more complex junctions, such as $Y$-junctions \cite{Y_junc_Buccheri,PRL_Ferraz_2019}, where we expect to see different behaviors for $W_{\te{ex}}(\tau)$ and $\lambda_{\te{opt}}(\tau)$, due to the presence of three-spin boundary interactions. As shown in the Appendix~\ref{apend} of this paper, as long as we only have a two-spin boundary interaction, the $M$-chain case is equivalent to the two-chain case.

\acknowledgements
M. F. C. would like to thank Rodrigo G. Pereira for useful comments about this manuscript and Isaac M. Carvalho for providing some preliminary numerical calculations related to this work. This work is supported by Conselho Nacional de Desenvolvimento Cient\'{i}fico e Tecnol\'{o}gico (CNPq), Brazil, through grant No. 200267/2023-0. M.V.S.B. acknowledges the support of CNPq under Grant No. 304120/2022-7 and Funda\c{c}\~ao de Amparo \`a Pesquisa do Estado de S\~ao Paulo (FAPESP) under Grant No. 2020/02170-4. EM acknowledges the support of CNPq (grant No. 309584/2021-3) and Fapesp (grant No. 22/15453-0). S.D. acknowledges support from  the U.S. National Science Foundation under Grant No. DMR-2010127 and the John Templeton Foundation under Grant No. 62422.

\appendix

\section{$\tau^{-2}$ scaling for the gapped situation}
\label{apend_B}

As discussed in the main text (see Eq. (\ref{relax_fun}) and Eq. (\ref{resp_fun_sit_I})), the relaxation function for situation I, where one chain is in the gapped phase and the other one in the gapless phase, is given by
\begin{equation}
    \Psi(t'-t)=-2g^2_0\sum_{k\leq k_F,k'}\frac{f^2_{kk'}}{\epsilon_{kk'}}\cos\left[\epsilon_{kk'}(t'-t)\right].
\end{equation}

In order to prove the universal $\tau^{-2}$ scaling of $W_{\te{ex}}(\tau)$ for large $\tau$, we will consider the general protocol given in Eq. (\ref{prot_expansion}), i.e., that $\dot \lambda(t/\tau)$ can be expanded in terms of the Chebyshev polynomials $T_{l}(2t/\tau-1)$. Combining this with the previous expression for the relaxation function, the double integral in Eq.~(\ref{excesswork}) can be written in terms of
\begin{equation}
    \int_{-1}^{1}dsds'\:T_{l}(s)T_{l'}(s')\cos\left[\epsilon_{kk'}\tau(s'-s)/2\right].
    \label{integration}
\end{equation}
As far as we know, this integration does not have an analytical solution for general $l,l'$. However, its large $\tau$ limit can be extracted easily. Since the product of the Chebyshev polynomials in the above equation ultimately contributes with $s^ns'^{m}$, where $n,m\geq 0$, the integration in Eq. (\ref{integration}) can be analyzed performing a sequence of multiple integration by parts. Each integration by parts produces a term $(\epsilon_{kk'}\tau)^{-1}$, so in the large $\tau$ limit, the leading contribution is  
\begin{equation}
    \int_{-1}^{1}dsds'\:T_{l}(s)T_{l'}(s')\cos\left[\epsilon_{kk'}\tau(s'-s)/2\right]\sim \frac{1}{(\epsilon_{kk'}\tau)^2},
    \label{integration_2}
\end{equation}
as long as $\epsilon_{kk'}$ never goes to zero (as indeed happens in this case, since $k\leq k_F$ and $k'\geq 0$). This 
shows that $W_{\te{ex}}(\tau\gg\Delta^{-1})\sim \tau^{-2}$ for \textit{any} protocol.

\section{$M$-chain case}
\label{apend}
In this Appendix, we show that as long as we have only two-spin boundary interactions, the $M$-chain case is equivalent to the two-chain case analyzed in the main text. 

Suppose  we want to make a weak junction between $M$ spin-$1/2$ chains. The system is described by the Hamiltonian
\begin{equation}
    \mc H(\lambda) = H_0 + \lambda(t)V,
\end{equation}
where  $H_0=\sum_{\ell=1}^{M}H_{\ell}$, with $H_\ell$ the Hamiltonian of chain $\ell$ [see Eq.~(\ref{H1})], and
\begin{equation}
    V=g_0\sum_{\ell=1}^{M}\left(S^{x}_{1,\ell}S^{x}_{1,\ell+1} + S^{y}_{1,\ell}S^{y}_{1,\ell+1} \right)\equiv \sum_{\ell=1}^{M}V_{\ell},
    \label{M_chain_boun_int}
\end{equation}
where $S^{a=x,y,z}_{1,\ell}$ is the spin operator at the first site of chain $\ell$ (with $M+\ell\equiv \ell$). Applying the generalized J-W transformation introduced in \cite{Nuclear_Cramp__2013}, this boundary interaction becomes [see Eq. (\ref{J_W_trans})] \cite{Nuclear_Giuliano_2020}:
\begin{equation}
    V_{\ell} = \frac{g_0}{2}\chi_{\ell}\chi_{\ell+1}\left[c^{\dagger}_{1,\ell}c_{1,\ell+1} -c^{\dagger}_{1,\ell+1}c_{1,\ell}  \right].
\end{equation}
The Klein factors $\chi_{\ell}=\chi^{\dagger}_{\ell}$ are responsible for keeping the correct anti-commutation relations, through $\{\chi_{\ell},\chi_{\ell'}\}=2\delta_{\ell\ell'}$, $\{\chi_{\ell},c_{j,\ell'}\}=\{\chi_{\ell},c^{\dagger}_{j,\ell'}\}=0$, between fermions in different chains.
From Eq.~(\ref{excesswork}), in order to determine  the excess work we need to evaluate the correlator $\langle V_I(t')V_I(t)\rangle_0$, now with $|\Upsilon_0\rangle=\prod_{\ell=1}^{M}|\Upsilon_{\ell,0}\rangle$, where $|\Upsilon_{\ell,0}\rangle$ is the ground state of chain $\ell$. Since the boundary interaction in Eq.~(\ref{M_chain_boun_int}) only couples pairs of chains and given the form of $|\Upsilon_0\rangle$, it is easy to obtain the result
\begin{eqnarray}
    \Phi(t'-t)&=&\sum_{\ell}\Phi_{\ell}(t'-t),
\end{eqnarray}
where $\Phi_{\ell}(t'-t)=i\langle [V_{I,\ell}(t'),V_{I,\ell}(t)]\rangle_0$ is the response function for each pair of chains. Thus, the relaxation function in Eq.~(\ref{relax_fun}) also assumes a decomposition similar to the above equation. Therefore, the excess work in Eq.~(\ref{excesswork}) is given by
\begin{equation}
    W_{\te{ex}}(\tau)=\sum_{\ell}W_{\te{ex},\ell}(\tau),
\end{equation}
where $W_{\te{ex},\ell}(\tau)$ is the excess work needed to establish a junction between only two chains.

\bibliography{references.bib}

\begin{thebibliography}{76}
\bibitem [{bos()}]{boso_correl}%
  \BibitemOpen
  \href@noop {} {}\bibinfo {note} {We use that $\langle e^{ia\phi_{\ell}(t)}e^{-ib\phi_{\ell'}(t')} \rangle_0=\frac{\delta_{ab}\delta_{\ell\ell'}}{[1+i\Lambda(t-t')]^{a^2/\pi K}}$ for $a \in \mathbbm R$.}\BibitemShut {Stop}%
\bibitem [{reg()}]{regularize_exp}%
  \BibitemOpen
  \href@noop {} {}\bibinfo {note} {Explicitly, they are given by $Y^l_{\mc M}=\frac{\mc M-l+1}{\mc M+1}\cos\left(\frac{\pi\: l}{\mc M+1}\right)+\frac{1}{\mc M+1}\sin\left(\frac{\pi\: l}{\mc M+1}\right)\cot\left(\frac{\pi}{\mc M+1}\right)$.}\BibitemShut {Stop}%
\end{thebibliography}

\end{document}